\documentclass[10pt]{article}
\usepackage{latexsym}
\usepackage{amsfonts}
\usepackage{graphicx}
\usepackage[numbers]{natbib}
\usepackage{subfig}
\usepackage{tikz}
\usepackage{multicol}
\usetikzlibrary{shapes.symbols,shapes.geometric,shadows,arrows.meta}
\tikzset{>={Latex[width=1.5mm,length=2mm]}}
\usepackage{flowchart}\usepackage[paperheight=11.69in,paperwidth=8.26in,left=0.61in,right=0.61in,top=0.61in,bottom=0.61in,headheight=1in]{geometry}
\usepackage[utf8]{inputenc}
\usepackage[T1]{fontenc}
\usepackage{helvet}

\usepackage[font=small,labelfont=bf]{caption}

\newenvironment{Figure}
  {\par\medskip\noindent\minipage{\linewidth}}
  {\endminipage\par\medskip}

\usepackage{tikz}
    \usetikzlibrary{shadows}
\usepackage{tcolorbox}
    \tcbuselibrary{skins}

\definecolor{esoblue}{HTML}{0070C0}
\newcommand{\mytitle}[1]{
    \node[fill=white,
        draw=esoblue,
        line width=0.5pt,
        text width=1.75cm,
        inner sep=8pt,
        xshift=-3.7cm]
    at (frame.north){\bfseries\textcolor{black}{#1}};
}

\newtcolorbox{mybox}[2][]{
    enhanced,
    overlay={\mytitle{#2}},
    borderline={.7pt}{0mm}{esoblue},
    frame hidden,
    arc=0mm,
    sidebyside,
    lefthand width=2.5cm,
    segmentation hidden,
    top=15pt,
    #1
}

\newcommand{\arttitle}[1]{\fontsize{24pt}{32pt}\selectfont \textcolor[HTML]{0070C0}{#1}\par}
\newcommand{\artauth}[1]{\fontsize{12pt}{18pt}\selectfont \textbf{#1}\par}
\newcommand{\artaff}[1]{\fontsize{11pt}{16pt}\selectfont #1 \par}
\newcommand{\TheTitle}{Improved strong lensing modelling of galaxy clusters using the Fundamental Plane: detailed mapping of the baryonic and dark matter mass distribution of Abell S1063}
\newcommand{\MyName}{Giovanni Granata}
\newcommand{\MyInst}{Dipartimento di Fisica, Universit\`a degli Studi di Milano, Via Celoria 16, I-20133 Milano, Italy}

\begin{document}

\begin{center}
\arttitle{\TheTitle}\par
\artauth{\MyName}\par
\artaff{\MyInst}
\end{center}\par

\addcontentsline{toc}{section}{{\MyName} - {\it \TheTitle}}
\vspace{0.5cm}
\begin{multicols}{2}
%%%%%%%%%%%%%%%%%%%% Here begins your scientific text %%%%%%%%%%%%%%%%%%%%

% use~\cite{XXX} for inserting references in text and provide citation at the end of the document

\noindent Galaxy clusters are the most massive gravitationally bound structures in the Universe, and around $85-90\%$ of their total mass is under the form of dark matter (DM). As a consequence, they are excellent astrophysical laboratories to test our hypoteses on the nature of DM itself. Thanks to several dedicated photometric and spectroscopic surveys, strong gravitational lensing (SL) has become the most accurate probe of the total mass distribution in the cores (out to a few hundreds of kiloparsecs from the centre) of massive galaxy clusters. SL can be combined with baryonic mass diagnostics to disentangle the mass distribution of cluster- and galaxy-scale DM haloes from the total mass distribution of the cluster. The resulting DM mass profiles can be compared to the predictions of high-resolution cosmological simulations, based on the $\Lambda$ cold dark matter (CDM) Cosmological Model.

The remarkable improvement in the accuracy of SL models, driven by recent observational campaigns, has allowed us to map robustly the mass distribution of the DM haloes hosting the member galaxies (usually referred to as sub-haloes). On this scale, a significant discrepancy between the predictions of SL models and high-resolution simulations has recently emerged: at a fixed galaxy total mass, sub-haloes extracted from SL models are more compact than their simulated counterparts~\cite{meneghetti20}.

The accuracy in the description of the cluster members in SL models is limited by the degeneracies between the parameters defining their mass distribution. These degeneracies have been reduced with the introduction of measured priors on the velocity dispersion of the cluster galaxies. However, the usual choice of adopting power-law scaling relations, with no scatter, to link the total mass of members with their luminosity is still a simplified approach. To obtain a more complex description of the cluster galaxies, we take advantage of measured kinematics and structural parameters of the cluster galaxies to calibrate the Fundamental Plane (FP) for the members of the massive galaxy cluster Abell S1063 (AS1063). Our results are presented in~\cite{granata22}

%%%% Start of a new section... Please change the section title
\vspace{0.25cm}
{\fontsize{10pt}{10.8pt}\selectfont \textcolor[HTML]{0070C0}{Building the strong lensing model of Abell S1063}\par}
\noindent AS1063 is one of the six massive clusters included in the Hubble Frontier Fields (HFF,~\cite{lotz17}) photometric survey. We build a SL model of its total mass distribution using the recent work by~\cite{bergamini19} as a starting point and reference. Like~\cite{bergamini19}, we model the diffuse DM and hot-gas mass distribution with isothermal cored haloes. The parameters of the hot-gas mass distribution are fixed from observations~\cite{bonamigo18}. The parameters of the cluster-scale DM mass distribution are optimised comparing the observed and model-predicted positions of the same multiply-imaged background sources. Likewise, we model the cluster galaxies with spherical, isothermal, truncated total mass distributions. The total mass profile of each member is entirely defined by two parameters: its velocity dispersion and its truncation radius.
While in~\cite{bergamini19} their values are derived with fixed power-law relations with respect to their observed total luminosity, we choose to consider the FP, a more accurate scaling law, which involves the magnitude, the half-light radius, and the central velocity dispersion of early-type galaxies.

We use HFF images in the F814W band to measure the values of the magnitude and of the half-light radius of all cluster members. Furthermore, for a sizeable subset of them, MUSE-VLT integral field spectroscopy allows us to determine the values of the line-of-sight central velocity dispersion. We calibrate the FP relation, and then use it to obtain the values of the velocity dispersion of all cluster members from their measured structural parameters. This procedure allows for a more accurate determination of the velocity dispersion values compared to the power-law approach. As for the truncation radii of the galaxies, we calibrate a proportionality relation with their observed half-light radii and use it to determine their values. Thanks to this procedure, we can thus fix the total mass distribution of all member galaxies in a new SL model of the cluster.

The more stringent constraints on the the mass distribution of the members compared to previous works result in a reduction of the statistical uncertainty on the parameters of the cluster-scale DM component, and in particular on the value of the core radius of the main DM halo of the cluster.
The new procedure also allows for more realistic scaling laws between the observables that describe the physical properties of the cluster members. For instance, the relation between the values of their total mass and their velocity dispersion is no longer a fixed power-law and now shows a significant scatter and a shallower slope.   

\vspace{0.25cm}
{\fontsize{10pt}{10.8pt}\selectfont \textcolor[HTML]{0070C0}{Mass profile decomposition}\par}
\noindent We decompose the total mass profile of the cluster derived from our new best-fit SL model into all its baryonic and DM components. As anticipated, the hot-gas mass profile is derived from X-ray observations. Combining instead the measured stellar mass values of the cluster members, presented in~\cite{mercurio21}, with their surface brightness profile in the HST F814W band, we derive the stellar mass profile of the cluster. We can thus disentangle the mass profiles of the various cluster- and galaxy-scale DM haloes by subtracting the baryonic component to the total mass distribution. This also allows us to derive the cumulative projected gas-, stellar-, and baryonic-to-total mass fractions out to a projected distance of $350 \, \mathrm{kpc}$ from the cluster centre: at this radius, we find a baryonic mass fraction of $0.147 \pm 0.002$. The profiles are presented in Figure~\ref{GranataFig1}.

%%% Example of a figure. Note the specific syntax!
\begin{Figure}
\centering
\includegraphics[width=\linewidth]{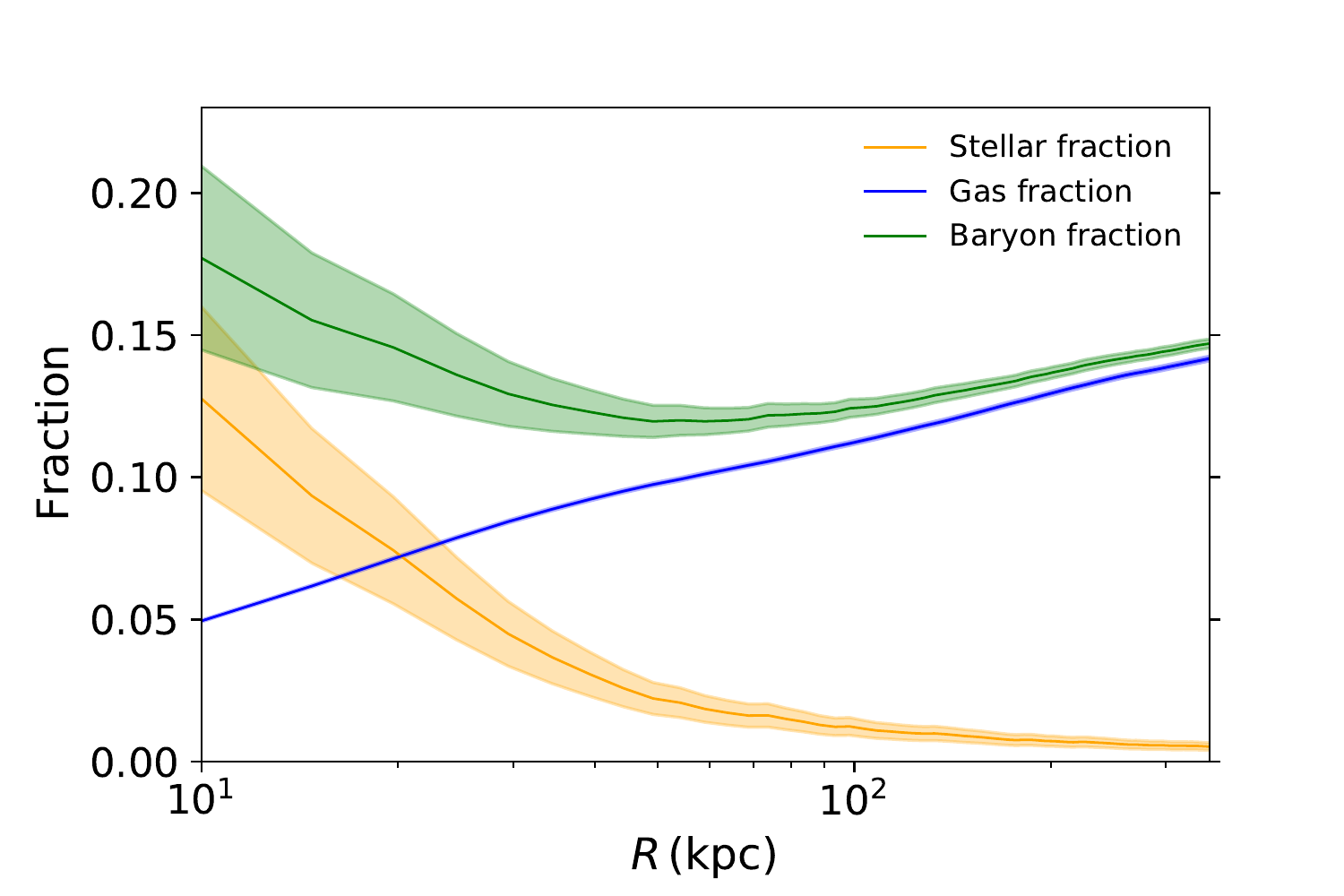}  %%% change name of figure
\captionof{figure}{\small Cumulative stellar-, gas-, and baryonic-to-total mass fractions from the model presented in~\cite{granata22}. Error bars are derived from the observations.}  %%% write caption
\label{GranataFig1}
\end{Figure}
\vspace{0.5cm}
%%% End of Figure

\vspace{0.25cm}
{\fontsize{10pt}{10.8pt}\selectfont \textcolor[HTML]{0070C0}{Comparison with cosmological simulations}\par}
\noindent As anticipated, comparing the physical properties of the DM sub-haloes as predicted by SL models to the most recent cosmological simulations is a test of the foundations of the $\Lambda$ CDM cosmological model on which the simulations are based, and of the micro-physics of DM. We first compare the stellar-to-total mass fractions of the cluster members with the predictions of recent HOD studies based on DM-only $N$-body simulations~\cite{girelli20}. We find a significant discrepancy: the stellar mass fraction values predicted by SL models are almost an order of magnitude higher than those predicted by the HOD procedure. This discrepancy is resolved if one considers, instead, hydrodynamical simulations, which include gas particles and stars, as well as the effects of the interaction between baryons and DM during the formation of clusters. We consider high-resolution simulations of clusters with a mass similar to that of AS1063 from~\cite{planelles14}. We perform two-dimensional projections to simulate the lensing observational conditions. In this case, we find compatible stellar mass fraction values from the SL model and the simulation suite.

Secondly, we examine how sub-haloes extracted from lensing models compare to their simulated counterparts in terms of maximum circular velocity, which is a proxy for their compactness.~\cite{meneghetti20} recently found that hydrodynamical simulations predict high-mass sub-haloes (total mass $M>10^{10} \, M_\odot$) to be significantly less compact than forecast by a sample of state-of-the-art SL models, including the model of Abell S1063 presented in~\cite{bergamini19}. The new technique we adopt significantly impacts the relation between the total mass and the maximum circular velocity of the sub-haloes, obtaining again a different slope compared to~\cite{bergamini19} and allowing for the inclusion of a scatter. However, as shown in Figure~\ref{GranataFig2}, our results agree with those from~\cite{bergamini19} in the mass range considered, thus confirming the reported discrepancy. Several tests to infer the origin of this discrepancy are being performed, focusing both on SL modelling and on the implementation of the cosmological simulations. However, no conclusive answer has been obtained so far. This leaves several open questions, and could point towards a new fundamental challenge for the $\Lambda$ CDM paradigm.     

%%% Example of a figure. Note the specific syntax!
\begin{Figure}
\centering
\includegraphics[width=\linewidth]{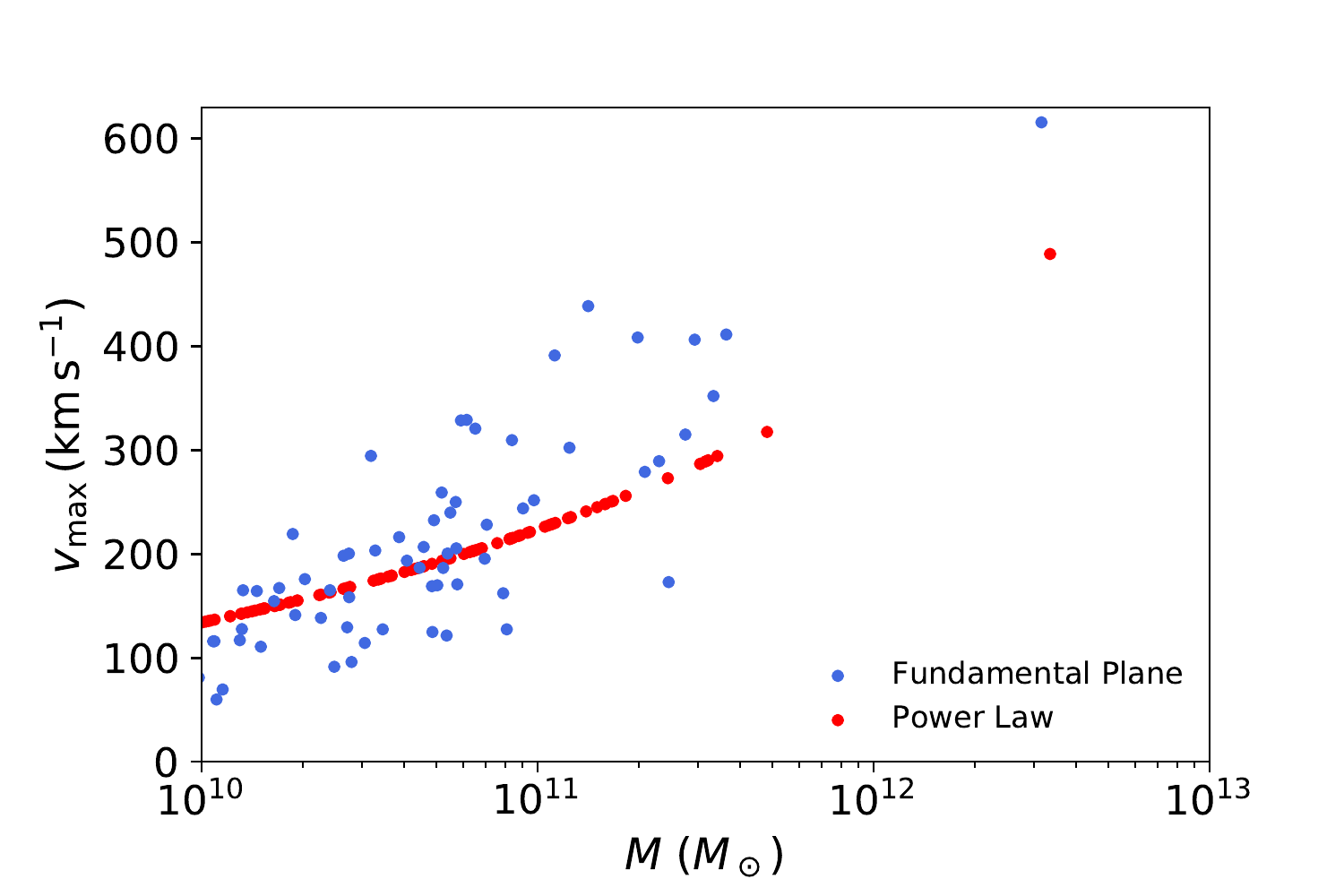}  %%% change name of figure
\captionof{figure}{\small Maximum circular velocity of the cluster members as a function of their total mass. The values predicted by the models presented in~\cite{granata22} and~\cite{bergamini19} are shown in blue and in red, respectively. Only members with $M>10^{10} \, M_{\odot}$ are shown.}  %%% write caption
\label{GranataFig2}
\end{Figure}
\vspace{0.5cm}
%%% End of Figure

%End here your main text

%%% List of References
\vspace{0.35cm}
%{\fontsize{10pt}{10.8pt}
{\fontsize{9pt}{9.8pt}
\selectfont \textcolor[HTML]{0070C0}{References}\par %}

\begingroup
\renewcommand{\section}[2]{}%

\endgroup
}
\end{multicols}

%%%% SHORT CV
\vspace*{\fill}
\begin{mybox}{Short CV}
    \includegraphics[scale=.06]{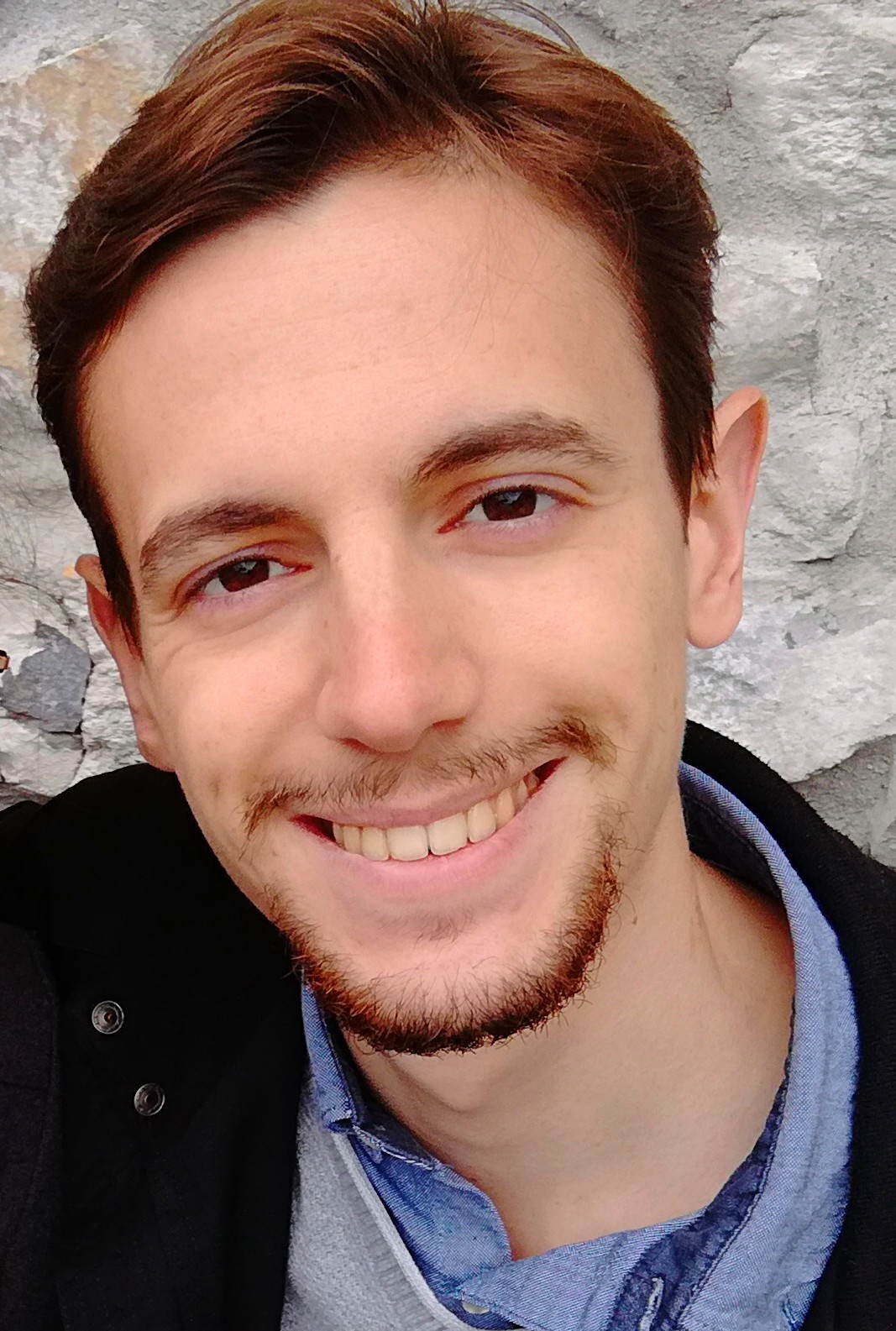}   %%% please put here your photo, which should be named as Name_Surname.jpg
    \tcblower
    %%% write the most important phases of your career
    \begin{tabular}{l@{\hspace{1\tabcolsep}}l}
    2018:& BSc in Physics, University of Milan\\
    2020:& MSc in Physics, University of Milan\\
    2020--present:& PhD in Physics, Astrophysics and applied Physics, University of Milan
    \end{tabular}
\end{mybox}

\end{document}